\documentclass{elsart}
\journal{Nuclear Physics A}
\usepackage{epsfig}
\usepackage{wasysym}
\usepackage{amssymb}
\usepackage{subfigure}
\usepackage{lscape}
\usepackage{supertabular}
\newcommand{\ind}{$^{115}$In }
\newcommand{\tin}{$^{115}$Sn }
\newcommand{\isome}{$^{115m}$In }

\begin{document}
\begin{frontmatter}

\title{Observation of $\beta$ decay of \ind to the first excited
level of \tin}
\author[a,b]{C. M. Cattadori},
\author[a]{M. De Deo},
\author[a]{M. Laubenstein},
\author[a,c]{L. Pandola\corauthref{cor}},
\ead{pandola@lngs.infn.it}
\corauth[cor]{INFN, Laboratori Nazionali del Gran Sasso,
S.S. 17 bis km 18+910, I-67010 L'Aquila, Italy.
Telephone number: +39 0862 437532, Fax number: +39 0862 437570}
\author[d]{V. I. Tretyak}
\address[a]{INFN, Laboratori Nazionali del Gran Sasso, S.S. 17 bis
km 18+910, I-67010 L'Aquila, Italy}
\address[b] {INFN Milano, Via Celoria 16, I-20133 Milano, Italy}
\address[c]{Universit\`a dell'Aquila, Dipartimento di Fisica, Via
Vetoio 1, I-67010 L'Aquila, Italy}
\address[d]{Institute for Nuclear Research, MSP 03680, Kiev, Ukraine}
\begin{abstract}
In the context of the LENS R\&D solar neutrino project, the
$\gamma$ spectrum of a sample of metallic indium was measured
using a single experimental setup of 4 HP-Ge detectors
located underground at the Gran Sasso National Laboratories (LNGS), Italy.
A $\gamma$ line at the energy (497.48$\pm$0.21) keV was found that is not 
present in
the background spectrum and that can be identified as a $\gamma$ quantum
following the $\beta$ decay of \ind to the first excited state of
$^{115}$Sn ($\frac{9}{2}^{+} \rightarrow \frac{3}{2}^{+}$).
This decay channel of $^{115}$In, which is reported here for the
first time, has an extremely low $Q_{\beta}$-value,
\mbox{$Q_{\beta}$ = (2$\pm$4)} keV, and has a much lower probability
than the
well-known ground state-ground state transition, being the
branching ratio $b = (1.18 \pm 0.31) \cdot 10^{-6}$.  
This could be the $\beta$ decay with the lowest known $Q_{\beta}$-value.  
The limit on charge non-conserving $\beta$ decay of $^{115}$In
is set at 90\% C.L. as $\tau_{CNC} > 4.1\cdot10^{20}$ y.
\end{abstract}
\end{frontmatter}
% main text
\section{Introduction}
\ind has been envisaged in the LENS project (Low Energy 
Neutrino Spectroscopy) as a possible target for the real-time measurement of 
low energy solar neutrinos. The detection principle is based on the inverse 
electron capture (EC) reaction $^{115}$In($\nu_{e}$,e$^{-}$)$^{115}$Sn$^{*}$ on 
$^{115}$In, which has a threshold energy of 114 keV and populates the 
second excited state of \tin at 613 keV (see Fig. \ref{levels} later). 
This state is metastable, with a life time $\tau$ = 4.76 $\mu$s, and its 
subsequent two-step decay to the ground state provides an highly specific 
$\nu$  signature \cite{LENS}. Namely, the prompt electron emitted in the 
inverse-EC 
reaction ($e_{1}$) is followed, with a typical delay of $\tau$=4.76 $\mu$s, 
by a localized deposition of 116 keV ($e/\gamma$)$_{2}$, in spatial 
coincidence, and  
by a $\gamma$-ray ($\gamma_{3}$) of 497 keV (see Fig. \ref{levels}).
Though \ind has several favorable features as a target 
for low energy solar neutrinos (high isotopic abundance, low threshold, 
strong $\nu$-tag), the detection technique is extremely challenging 
because \ind is unstable and can $\beta$-decay directly to the 
\tin ground state. The specific activity of natural indium is 0.25 
Bq/g and thus indium itself is the major irremovable source of background. 
The LENS R\&D project has demonstrated that the 10$^{6}$ background 
suppression factor needed for the $\emph{pp}$ solar neutrino measurement 
can be achieved, however at the cost of a very high segmentation 
($\sim 10^{5}$ cells) and of an one-order-of-magnitude increase of the 
overall detector mass \cite{LENS}. \\
The most serious background that has to be faced is related to the 
coincidence of two spatially-close indium decays. The coincidence of two 
\ind decays can mimic the $\nu$-tag if the second $\beta$ has energy 
close to the end point and emits a hard bremsstrahlung $\gamma$. The 
importance of this source of background depends on the granularity of the 
detector (because of the requirement of spatial coincidence), on the 
energy resolution and on the indium bremsstrahlung spectrum. In order to 
have a better comprehension and characterization of the bremsstrahlung, a 
measurement of the $\gamma$ spectrum of 
an indium sample was performed  with HP-Ge detectors 
in the Low-Level Background Facility underground in the Gran Sasso 
National Laboratories, Italy. During the analysis of the data we found evidence 
for a previously unknown decay of \ind to the first excited level 
of $^{115}$Sn, at the energy 497.4 keV. \\
In Section \ref{sectuno} we describe the indium sample and the experimental 
details of the measurement. In Section \ref{sectdue} we present the spectrum and 
the evidence for the $\gamma$ line at 497 keV. In Section \ref{secttre} we 
discuss the interpretation of this line in terms of decay of \ind to the 
first excited state of \tin and the consequent limit set on the \ind 
charge non-conserving decay. Finally, in Section \ref{sectquattro} we 
summarize our results and briefly describe some 
possible future perspectives.
\section{The setup} \label{sectuno}
\subsection{The sample and the detector}
\begin{figure}[t]
\begin{center}
\mbox{\epsfig{file=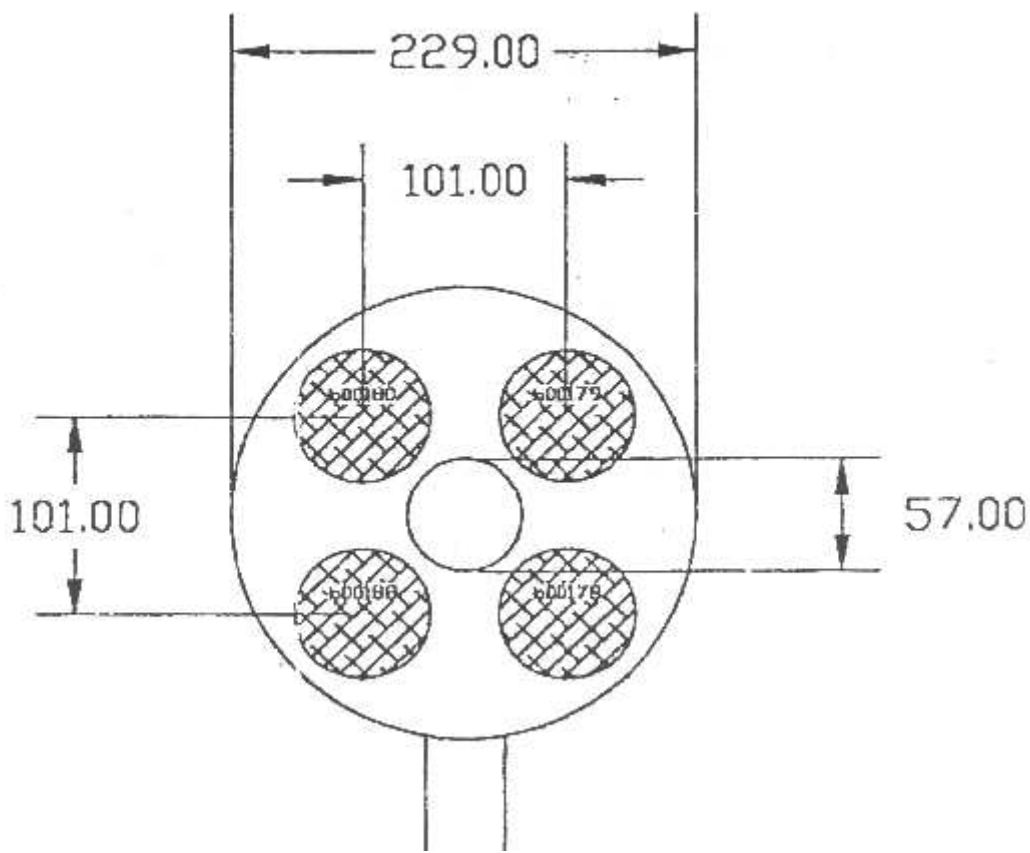,height=12cm}}
\caption{Top view of the experimental setup. Dimensions are in mm.}
  \label{setupplot}
\end{center}
\end{figure}

The sample used for the measurement consists of an ingot of metallic $^{nat}$In
of mass (928.7$\pm$0.1) g. It has the shape of a cylindrical shell, with the 
approximate
dimensions of 2.0 cm (internal diameter), 5.5 cm (external diameter) and 6.5 cm
(height).
The high-purity indium (6N5 grade) used for the production of the sample has been
provided by the Indium Corporation of America in May 2002. \\
The $\gamma$ spectrum of the indium source was measured using a set of 4 HP-Ge detectors
installed underground
at LNGS. The detectors are very similar coaxial
germanium crystals mounted altogether in one cryostat made by
Canberra; their main parameters are displayed in Table
\ref{detectors}. They are arranged as shown in Fig. \ref{setupplot}
and surround the indium sample, which is placed in the central well.
The experimental setup is enclosed in
a lead/copper passive shielding and has a nitrogen ventilation system against
radon. \\
One measurement of the indium ingot (2762.3 h of counting time for
each detector) and one of the background (1601.0 h) were carried out, 
both with the complete shielding around the detectors.
\nopagebreak
\begin{table}[t]
\caption{Germanium detectors parameters} \label{detectors}
\begin{center}
\begin{tabular}{|l|c c c c|}
\hline
  & \multicolumn{4}{|c|}{         Detectors         } \\
  & ge178 & ge179 & ge180 & ge188 \\
\hline
Volume (cm$^{3}$) & 225.2 & 225.0 & 225.0 & 220.7 \\
Endcap and holder material & \multicolumn{4}{|c|}{Electrolytical Copper} \\
Energy resolution (FWHM) at 1332 keV & 2.1 & 2.0 & 2.0 & 2.0 \\
\hline
\end{tabular}
\end{center}
\end{table}
\subsection{Detector performances} \label{MC}
\begin{figure}[t]
\begin{center}
\mbox{\epsfig{file=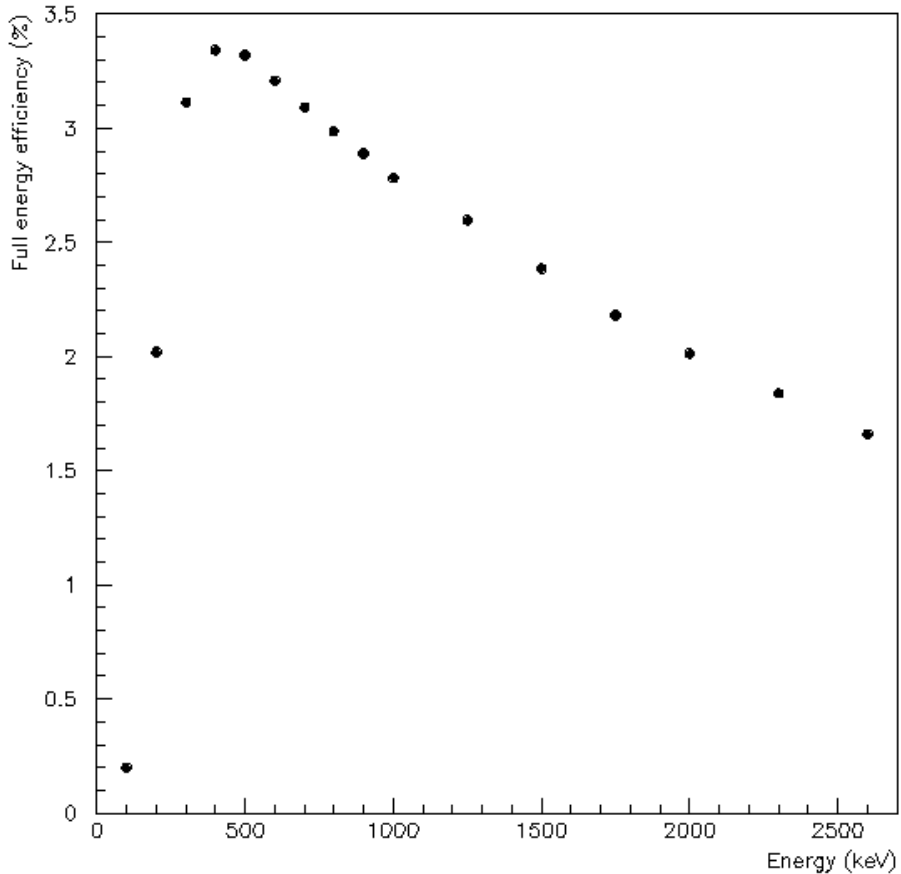,height=12cm}}
\caption{Simulated full energy efficiency of the 4-detector setup}
  \label{effplot}
\end{center}
\end{figure}

The detectors show a very good energy resolution and linearity.
The efficiency of the
configuration (detectors and ingot) has been evaluated using the
\textsc{Geant4}-based code \textsc{Jazzy}, developed by O.~Cremonesi\footnote{INFN Milano 
and Milano Bicocca University.}. The computed full
energy efficiency, i.e. the probability that a $\gamma$ ray produced in a random position
of the indium ingot deposits its full energy in one of the four Ge detectors, is displayed
in Fig. \ref{effplot} as a function of energy.
\nopagebreak
The Monte Carlo code has been checked, with the
procedure described in Ref. \cite{neodimio}, using a previous measurement performed in
the same setup with a $^{60}$Co $\gamma$ source (1173 keV and 1332 keV).
Taking into account the 2505 keV sum peak in the
experimental spectrum and neglecting the angular correlations of the two photons, the 
measured absolute efficiencies in this configuration agree with the
computed ones at 12$\%$ and are consistent within their statistical uncertainities.
On the basis of our previous experience with similar simulations of analogous experimental
setups we estimate the systematic uncertainty on the Monte
Carlo efficiencies to be 10\%.
In any case, the knowledge of the absolute efficiency at the
few $\%$ level is not needed for the present analysis, as it does not represent the main
contribution to the uncertainty on the measured \ind $\rightarrow$ $^{115}$Sn$^{*}$ decay rate
(see Sect. \ref{secttre}). For this reason, we can conclude that our knowledge of the
detector and of the absolute efficiency is satisfactory for our purposes.
\section{Analysis of the measured spectra} \label{sectdue}
\subsection{Comparison of indium and background spectra} \label{spectrum_des}
\nopagebreak
\begin{figure}[p]
\begin{center}
\mbox{\epsfig{file=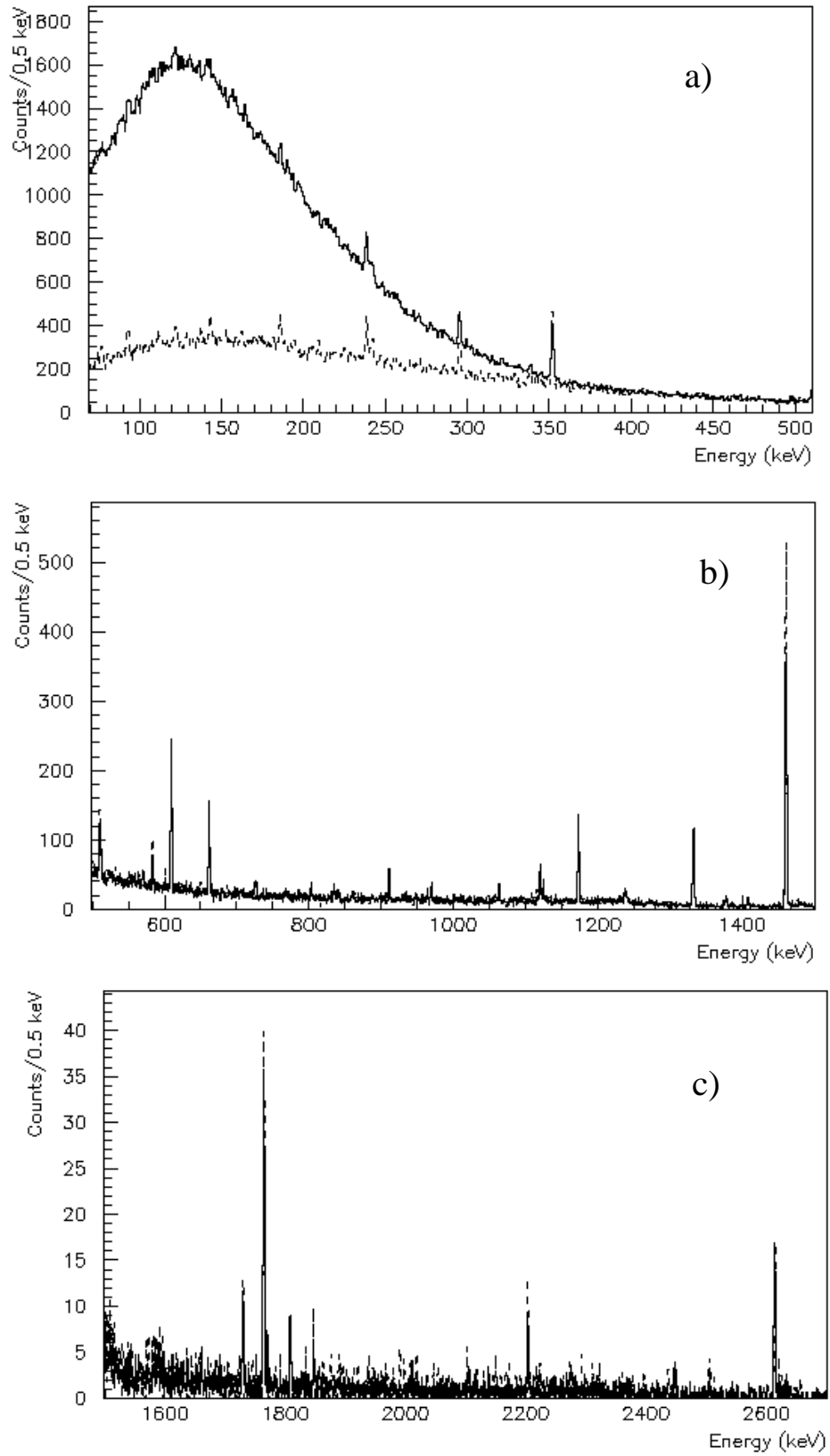,width=13cm}}
\caption{Experimental $\gamma$ spectrum in the regions 70$-$510 keV (a), 500$-$1500 keV (b) and
1500$-$2700 keV (c) for the indium sample (solid line) superimposed with the background spectrum
(dashed line) normalized to the same counting time.} \label{brem}
\end{center}
\end{figure}
The spectra obtained by the four detectors for the indium sample and for the background
have been added (see Fig. \ref{brem}) and analyzed using both
an automatic program
and visual inspection. In the
indium measurement, besides the continuous component (see Fig. \ref{brem}a) due to the bremsstrahlung of
the electrons from the $\beta$ decay of \ind (end point: 499 keV), 42 $\gamma$ lines
with energy above 200 keV could be identified: they are listed, together with their interpretation,
in Table \ref{gammatable}. The same $\gamma$ lines, except the one at (497.48$\pm$0.21)
keV (which represents the main point of this paper and will be discussed in detail in the
following sections), are also found in the background spectrum, though some of them (e.g. 
795, 1588 and 2447 keV) have a poor statistical significance because of the shorter
measurement time. As shown in Table \ref{gammatable}, the observed $\gamma$ lines come from the
natural radionuclides and radioactive series ($^{40}$K, $^{238}$U, $^{235}$U, $^{232}$Th) and
from cosmogenic or antropogenic nuclides ($^{60}$Co, $^{137}$Cs, $^{207}$Bi,
$^{26}$Al) that are usually
present as contaminations in normal copper and lead. The same Table
shows the counting rates of the
$\gamma$ lines for the indium sample and the background, as well as their difference
(statistical errors only). For each line (except the one at 497 keV) the difference turns out to be
statistically consistent with zero. Hence there is 
no statistical evidence of radioactive contamination of the indium sample, since the data are
consistent with contaminations of the experimental setup only (germanium crystals and passive
lead/copper shielding).\\
The efficiencies quoted in Table
\ref{gammatable} (full energy efficiencies) have been calculated in the hypothesis that the
$\gamma$ rays are generated inside the indium sample so they include the effect of 
self-absorption in the ingot itself. 
The fact that in some cases the indium-background
difference rate is negative (though statistically consistent with zero) is explained because
indium is an effective $\gamma$ ray absorber (the atomic number $Z$ is 49) and can hence act as an
additional shielding for the Ge detectors with respect to the background measurement, where the 
well is filled with a plexiglas plug.
\subsection{The 497 keV line} \label{luciano}
\begin{figure}[t]
\begin{center}
\mbox{\epsfig{file=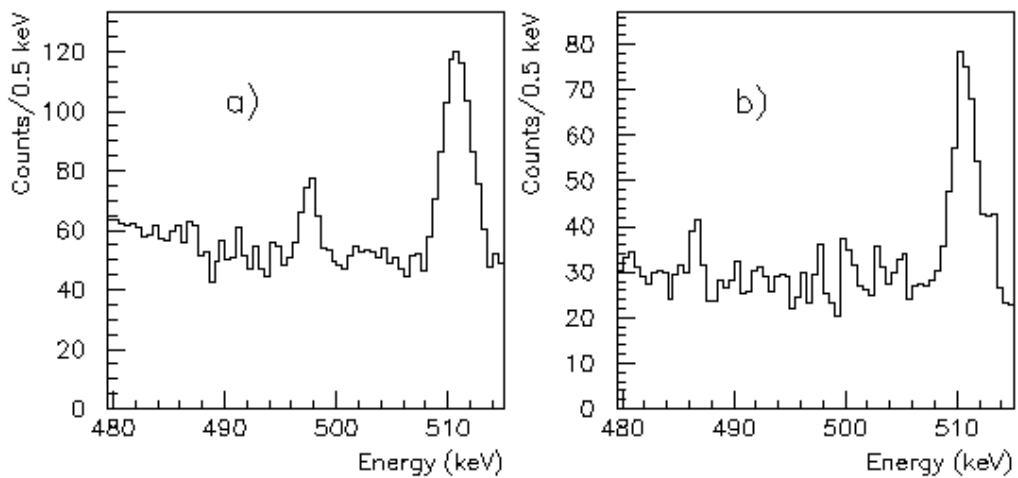,width=14cm}}
\caption{Spectrum in the region 480$-$515 keV (a) for the indium sample and (b) for the background.}
  \label{lineplot}
\end{center}
\end{figure}

As anticipated in Sect. \ref{spectrum_des}, the only $\gamma$ line of the indium
spectrum which is not present in the background measurement and cannot be ascribed
to the usual radioactive contaminants is located at the energy of (497.48$\pm$0.21) keV. 
The interesting region of the spectrum is shown in Fig. \ref{lineplot}.
\nopagebreak
From the fit of the indium spectrum in the
energy region (487-508) keV with a Gaussian peak and linear background assumption, 
we get a net area of (90$\pm$22) counts, inconsistent
with zero at more than 4$\sigma$. This
corresponds to a counting rate of (0.78$\pm$0.19) counts/day. Variations
of the energy interval for the fit result in changes of the area %from 84 to 103 counts,
inside the quoted uncertainty.
With the same procedure applied to the background spectrum, no Gaussian peak could be found
and the resulting area is (0$\pm$14) counts; the corresponding upper limit derived with the 
Feldman-Cousins method \cite{feld} is 23.0 counts (90$\%$ CL) or 
0.34 counts/day (90$\%$ CL). \\
It can hence be concluded that the peak under examination is statistically significant and
related with the indium sample, being absent in the background measurement.

\section{Interpretation} \label{secttre}
\subsection{Decay of $^{115}$\textrm{In} to the first excited state of $^{115}$\mbox{Sn}}
\label{betaexc}
\begin{figure}[t]
\begin{center}
\mbox{\epsfig{file=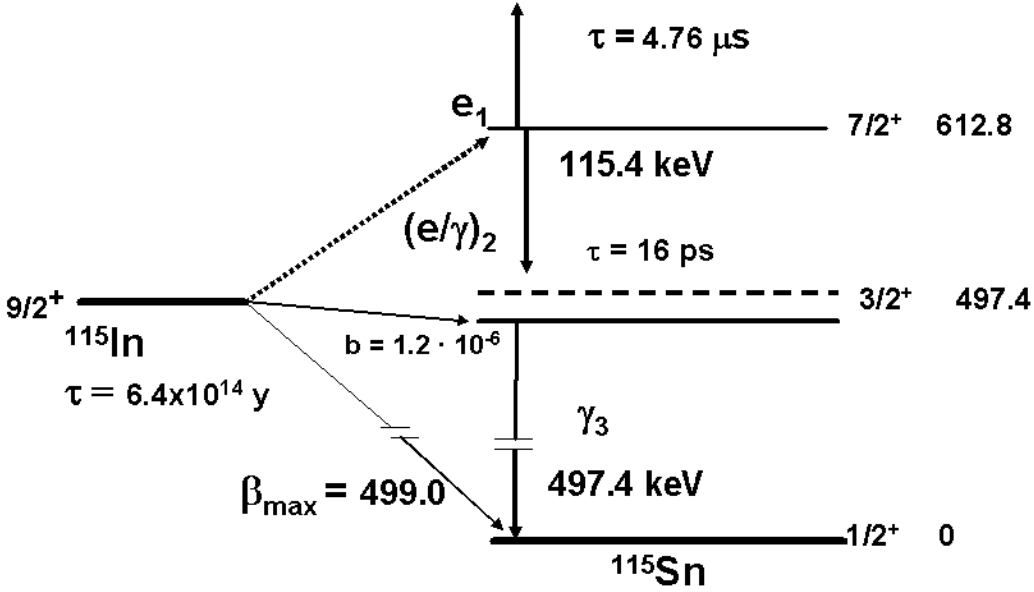,width=14cm}}
\caption{New level scheme for \ind and $^{115}$Sn as a result after this work. The $\frac{3}{2}^{+}$ 
level of \tin is shown 
in its old (dashed line) and new (solid line) position, namely above and below the \ind ground state.
The nuclear transitions relevant for the LENS
neutrino tagging are also shown.} \label{levels}
\end{center}
\end{figure}

The peak with the energy (497.48$\pm$0.21) keV found in the indium measurement and absent in
the background one can be explained with the $\beta$ decay of \ind to the first excited
level of $^{115}$Sn, whose excitation energy is 497.4 keV. Such a process has never been
observed previously and the $\beta$ decay of \ind was considered
up-to-date as going exclusively to the ground state of \tin \cite{isotopes,blachot,audi1}.
Because of the large change in the nuclear angular momentum
($\frac{9}{2}^{+} \rightarrow \frac{1}{2}^{+}$),
the ground state to ground state $\beta$ decay of \ind is a 4$^{th}$ forbidden transition
and has one of the largest known $\log ft$ values ($\log ft =$ 22.5). The measured half life
of \ind is $t_{1/2} = 4.41 \cdot 10^{14}$ y \cite{isotopes,blachot,audi1,Pfe79}. \\
According to the most recent table of atomic masses \cite{audi2}, the mass difference
between \ind and \tin is (499$\pm$4) keV.
The decay \ind $\rightarrow ^{115}$Sn$^{*}$ to the first excited level of \tin is hence
kinematically allowed, though with an extremely small $Q_{\beta}$ value,
\mbox{$Q_{\beta}$ = (1.6$\pm$4.0)} keV. \\
Using the area of the 497 keV peak observed in the indium spectrum, the
decay rate for the transition to the first excited level of \tin can be evaluated 
through the relation
\begin{equation}
\Gamma (^{115}\mbox{In} \rightarrow ^{115}\mbox{Sn}^{*}) \ = \ \frac{S (1+\alpha)}{N \
\varepsilon \ t_{M}},
\end{equation}
where
$\varepsilon$ is the efficiency to detect the full energy $\gamma$ with the 4 Ge
detectors, $N$ is the number of \ind nuclei in the sample, $t_{M}$ is the measurement time, $S$ is
the area of the peak and $\alpha$ is the coefficient of conversion of $\gamma$ quanta to electrons
for the given nuclear transition. \\
The full peak efficiency at 497 keV is $\varepsilon$ = (3.32$\pm$0.33)\% and it was evaluated
using the Monte Carlo simulation described in Sect. \ref{MC}.
Taking into account the total mass of the indium sample (928.7 g), the atomic weight of indium
(114.8 g$\cdot$mol$^{-1}$) \cite{weights} and
the isotopic abundance of \ind (95.7\%) \cite{chem}, the number of \ind nuclei in our sample
results to be $N$=4.66$\cdot 10^{24}$.
The area of the peak is (90$\pm$22) counts (see Sect. \ref{luciano}) and the electron
conversion coefficient for the transition is $\alpha$=8.1$\cdot 10^{-3}$ \cite{blachot}.
With these values and $t_{M}$=2762.3 h, the decay rate for the \ind $\beta$
decay to the first excited level of $^{115}$Sn is estimated to be
\begin{equation}
\Gamma (^{115}\mbox{In} \rightarrow ^{115}\mbox{Sn}^{*}) \ =
\ (1.86 \pm 0.49) \cdot 10^{-21} \ \ \textrm{y}^{-1},
\end{equation}
that corresponds to a partial half life of
\begin{equation}
t_{1/2} (^{115}\mbox{In} \rightarrow ^{115}\mbox{Sn}^{*}) \ =
\  \frac{\ln 2}{\Gamma} \ = \  (3.73 \pm 0.98) \cdot 10^{20} \ \ \textrm{y}.
\end{equation}
The probability of this process is thus near one million times lower than for the transition to 
the ground state of \tin (see Fig. \ref{levels}); the
experimental branching ratio is
\begin{equation}
b \ = \ (1.18 \pm 0.31) \cdot 10^{-6}.
\end{equation}
The uncertainty on the decay rate and on the branching ratio mainly comes from the statistical error 
on the net area of the 497 keV peak.
\nopagebreak
Nuclear spin and parity are changed in the observed transition  
from the initial $\frac{9}{2}^{+}$ of the
\ind ground state to $\frac{3}{2}^{+}$ of
$^{115}$Sn$^*$; this is therefore a 2$^{nd}$ forbidden unique $\beta$ decay.
The recommended value of $\log ft$ for such a decay is equal to
(11.7$\pm$0.9) \cite{Akh88}. With the measured value of the half life 
of $(3.73 \pm 0.98) \cdot 10^{20}$ y,
the ``experimental'' $\log f$ value can be calculated as 
$\log f = (-16.37 \pm 0.91)$.
On the other hand, the $\log f$ value can be estimated using
the formalism described in \cite{Dzh72}:
\begin{equation}
f = f_0 \ s_2,
\end{equation}
\noindent
where $f_0$ is the value for the allowed transition and $s_2$ is
the correction for the 2$^{nd}$ forbidden unique decay.
The $f_0$ factor can be written as 
\begin{equation}
f_0 = \phi_0 s_0,
\end{equation}
where $\phi_0$
is the phase space integral
\begin{equation}
\phi_0 = \int_1^{W_0} W \sqrt{W^2-1} (W_0-W)^2 dW,
\end{equation}
$W_0$ is the energy release in the $\beta$-decay in units of the electron mass 
and $s_0$ is the correction for a non-zero $Z$ value of the daughter nucleus. The
phase space integral can be analytically calculated as
\begin{equation}
\phi_0 = \sqrt{W_0^2-1}\frac{2W_0^4-9W_0^2-8}{60} + \frac{W_0}{4} 
\ln \Big( W_0+\sqrt{W_0^2-1} \Big),
\end{equation}
while the $s_0$ values are tabulated in \cite{Dzh72}.
The correction $s_2$ can be estimated following Ref. \cite{Dav51} as:
\begin{equation}
s_2 = \frac{1}{1080} 
\Big[ \frac{3}{7}(W_0^2-1)^2 - \frac{26}{105}(W_0^2-1)(W_0-1)
- \frac{2}{105}(W_0-1)^2 \Big].
\end{equation}
In this way the $\log f$ value for $Q_\beta=1.6$ keV is equal to
$-15.78$, which is in agreement with the experimental value of
$\log f = (-16.37 \pm 0.91)$. \\
The $\log f$ value can also be evaluated with the help of the LOGFT tool
at the National Nuclear Data Center, USA \cite{logft} which is based on
procedures described in \cite{Gov71}. The 
value of $\log f$ for $Q_\beta=1.6$ keV calculated with the LOGFT code 
is however very different from the
previous estimation: $\log f = -10.8$; 
this means that with such a $Q_\beta$ the beta decay should go near 6 orders of
magnitude faster. One can solve the inverse problem and use the LOGFT code to
adjust the $Q_\beta$ value corresponding to the measured 
$\log f = (-16.37 \pm 0.91)$.
Such a procedure gives an extremely low value of
$Q_\beta = (23^{+23}_{-12})$ eV. \\
While both these calculations can be considered as very tentative and 
not being
intended to be used for such low energies, the last estimation gives a hint
that the $Q_\beta$ value in the beta decay $^{115}$In $\rightarrow$ $^{115}$Sn$^*$
can be very close to zero\footnote{Even the history of the $Q_\beta$ evaluation
for $^{115}$In gives some indication for this:
the $Q_\beta$ value was slightly
lower than energy of the first excited 497.4 keV $^{115}$Sn state
in accordance with older tables of atomic masses, 
$Q_\beta=(495 \pm 4)$ keV \cite{Aud93} and ($496 \pm 4$) keV \cite{Aud95}, 
while it is slightly higher in the last evaluation, ($499 \pm 4$) keV \cite{audi2}.}.
Such a unique situation could be used to establish a limit on the
antineutrino mass,
in addition to the experiments with $^{3}$H and $^{187}$Re, where up-to-date limits
are in the range of $\simeq$ 2 eV \cite{Lob03} and $\simeq$ 15 eV \cite{Sis04}, respectively.
To do this in a competitive way both the \ind $Q_\beta$-value and the energy
of the \tin 497.4 keV level should be measured with an accuracy of $\sim$1 eV
or better.
The uncertainty in the energy of the 497.4 keV level is equal now to 22 eV \cite{blachot}
and it could be further reduced doing an accurate investigation of the $^{115}$Sb 
decay\footnote{It can be noted that energies of many $\gamma$ lines of calibration sources
are known with an accuracy of $(0.1-0.3)$ eV, also in the $\sim$500 keV region
of our interest \cite{isotopes}.}.
The atomic mass difference \ind -- \tin can possibly also be measured with an 
accuracy of  $\sim$1 eV. For example, the mass difference of $^{76}$Ge --
$^{76}$Se was determined with 50 eV uncertainty in Ref. \cite{Dou01}, 
where it was stated that the Penning traps technique is able to deliver
even more accurate results.
Both such measurements require strong experimental efforts but the 
physical result could be very interesting and important.
\subsection{Possible imitation of the effect}
In some nuclear processes $\gamma$ rays with energies close to 497 keV are emitted. This
could give an alternative explanation of the peak observed in the experimental spectrum. Luckily,
additional $\gamma$ rays are also emitted in such decays, allowing to tag those
mimicking effects.\\
The \ind nucleus has an isomeric state \isome with the energy $E_{iso}$ = 336.2 keV and 
a half life of 4.5 h \cite{isotopes}. With the probability of 0.047\% 
the \isome nucleus $\beta$-decays
to the first excited level of $^{115}$Sn, with the subsequent emission of a 497 keV $\gamma$ ray
\cite{isotopes,blachot,audi1}.
However, in this case a $\gamma$ ray with the energy $E_{iso}$ = 336.2 keV is emitted
with much higher probability (45.84$\%$ \cite{isotopes}) because of the electromagnetic transition
from the isomeric \isome to the ground \ind state.
This huge peak at 336.2 keV, whose area should be $\sim$ 10$^{3}$
times bigger than that of the observed 497.4 keV peak, is absent in the
experimental spectrum; only a peak at 338.3 keV is observed, with the net area of 
(138$\pm$50) counts,
which corresponds to the decay of $^{228}$Ac from the $^{232}$Th natural chain
(see Table \ref{gammatable}).
Therefore the decay of the isomeric state \isome is absolutely negligible
and the 497 keV peak cannot be ascribed to it, not even in part.
Similarly, given also the underground location of the
experimental setup and the low flux of neutrons \cite{belli}, 
we conclude that $(n,\gamma)$ reactions cannot
contribute to the peak under analysis.\\
Protons produced by fast neutron or cosmic ray muons can populate the second excited level
of \tin (Fig. \ref{levels}) via the $(p,n)$ reaction on \ind ($E_{thr}$=0.9 MeV);
the \tin nucleus quickly returns to the
ground state with the emission of two $\gamma$ rays of energy 115.4 and 497.4 keV. The contribution
originated by fast neutrons is practically zero (see f.i. \cite{gallex1}) because of the deep
underground location and the lack of hydrogenous materials in the setup.  On the 
other hand, since the
muon flux in the laboratory is extremely low (1 $\mu$/m$^{2}$h \cite{macro}),
also the contribution
of $(p,n)$ reactions induced by cosmic rays (see also \cite{gallex2}) to the 497 keV peak
is absolutely negligible ($<$ 10$^{-3}$ counts). \\
Some decays from the natural $^{238}$U and $^{232}$Th chains can also give $\gamma$ rays
in the energy region of interest, though with very low intensity.
They are in particular \cite{isotopes} $^{214}$Bi ($E$ = 496.7 keV, $I$ = 0.0069\%), $^{228}$Ac
($E$ = 497.5 keV, $I$ = 0.0059\%) and $^{234m}$Pa ($E$ = 498.0 keV, $I$ = 0.062\%).
However, the sum
contribution of these decays to the 497 keV peak is less than 1 count and can be easily estimated
using their stronger associated $\gamma$ lines.\footnote{For instance, the area of the 338.3 keV line
of $^{228}$Ac, whose relative intensity is 11.27\%, is only (138$\pm$50) counts. Therefore, if the
contamination were located in the indium ingot, the estimated contribution to the 497 keV peak,
taking also into account the different full peak efficiency, would be (7.3$\pm$2.6)$\cdot$10$^{-2}$
counts.}\\
We could not figure out other sources than can mimic the experimentally observed 497 keV peak.
\subsection{Charge non-conserving $\beta$ decay of \ind}
The present measurements give also the possibility to set a limit on 
the charge non-conserving
(CNC) $\beta$ decay of $^{115}$In,
a process in which the $(A,Z) \to (A,Z+1)$ transformation is not accompanied
by the emission of an electron \cite{Fei59,Oku92}. It is supposed that instead of
an $e^-$, a massless particle is emitted (for example a $\nu_e$, a $\gamma$
quantum, or a Majoron):
$(A,Z) \to (A,Z+1) + (\nu_e$ or $\gamma$ or M) + $\overline{\nu}_e$.
Up-to-date, the CNC $\beta$ decay was searched for with four nuclides only:
$^{71}$Ga, $^{73}$Ge, $^{87}$Rb and $^{113}$Cd. 
Only lower limits on corresponding life times $\tau_{CNC}$ were
established, in the range of ($10^{18}-10^{26}$) y (we refer to the last work
on this subject \cite{Kli02}, where all previous attempts are reviewed).
Measuring only the $\gamma$ quanta from the deexcitation of the 497 keV level
of $^{115}$Sn, we cannot distinguish, in principle,
which mechanism leads to the detected 497 keV peak: the CNC beta decay or
the usual charge-conserving beta decay to the excited level of $^{115}$Sn.
If the whole area of the observed 497 keV peak is considered as belonging 
to the CNC beta decay
(instead of the much less exotic usual $\beta$-decay of $^{115}$In),
and substituting in the formula for the life time
$\tau_{CNC} = \varepsilon \cdot N \cdot t_M / [S_{lim}(1+\alpha)]$
the values of efficiency and other parameters described in section \ref{betaexc}
together with $S_{lim}=118$, the Feldman-Cousins 90\% CL limit on the 
life time of this process is
\begin{equation}
\tau_{CNC} \ > \ 4.1 \cdot 10^{20} \ \textrm{y}.
\end{equation}
Though this value is relatively low, it is determined for the first time for
$^{115}$In, expanding the scarce information on the CNC processes.
\section{Conclusions} \label{sectquattro}
From the measurement of the $\gamma$ spectrum of a sample of metallic indium 
performed with HP-Ge detectors in the Low-Level Background Facility of the 
Gran Sasso 
Laboratory, we have found evidence for the previously unknown $\beta$ decay 
of \ind to the first excited state of $^{115}$Sn at 497.4 keV 
($\frac{9}{2}^{+} \rightarrow \frac{3}{2}^{+}$). 
The $Q_{\beta}$-value for this channel is {$Q_{\beta}$ = (2$\pm$4)} keV; this 
could be lower than the $Q_{\beta}$-value of $^{163}$Ho, 
$Q_{\beta}$=2.565 keV \cite{isotopes,olmio}, 
and hence be the lowest of all the known $\beta$ decays. The branching 
ratio is found to be $b = (1.18 \pm 0.31) \cdot 10^{-6}$. We also set a 
limit on the charge non-conserving $\beta$ decay of \ind 
$\tau_{CNC} > 4.1\cdot10^{20}$ y (90\% CL). \\
This measurement was carried out in the context of the LENS R\&D project, 
with the aim of better characterize the \ind bremsstrahlung spectrum, which is 
poorly known near its end-point.
The discovered decay of \ind on the first excited state of \tin could in 
principle be dangerous for the LENS solar neutrino measurement, since it 
is an irremovable background source of 497 keV $\gamma$ rays.   
The coincidence of \ind decay on the ground state of \tin and one \ind decay 
on the first excited state can in fact mimic the $\nu$-tag and  
the possible impact is currently under investigation.\\
In a future work we plan to study the possible atomic effects on the half life of the 
\ind $\rightarrow$ $^{115}$Sn$^{*}$ decay by measuring a new sample where indium is 
present in a different chemical form (e.g. InCl$_{3}$ solution instead of metallic 
indium). \\
We also point out that, given the extremely low $Q_{\beta}$ value, the decay 
reported in this work could be used in a future experiment to directly measure the 
neutrino mass. In order to reduce the background due to the $\beta$ decay of \ind 
to the ground state of $^{115}$Sn, such experiment would need a rejection power 
of $10^{6}$, that could be achieved tagging the 497 keV $\gamma$ ray emitted in 
coincidence with the $\beta$ particle. New 
In-based semiconductor detectors or fast bolometers could be used for the purpose. 
\section{Acknowledgments}
The authors would like to thank O.~Cremonesi, for providing his Monte Carlo code to estimate the
efficiencies, E.~Bellotti and Yu.G.~Zdesenko, for their continuous support and useful advices, 
and A.~di Vacri, who
collaborated in the early stages of this work. We would also like to express our gratitude to
D.~Motta, for his valuable advices and feedback, and to the LNGS Mechanical Workshop, especially 
to E.~Tatananni and B.~Romualdi, for the melting of the indium ingot.
\begin{landscape}
\begin{center}
\tablefirsthead{%
\hline
\multicolumn{4}{|c|}{Database lines} & & \multicolumn{2}{|c|}{Indium spectrum} & \multicolumn{2}{|c|}{Background spectrum} 
& Difference \\
Energy & Decay & Chain & Intensity & $\varepsilon$ & Position & Rate & Position & Rate & \\
(keV) &  &  & ($\%$) & ($\%$) & (keV) &  c/day & (keV) & c/day & c/day \\
\hline}
\tablehead{
\hline
\multicolumn{4}{|c|}{Database lines} & & \multicolumn{2}{|c|}{Indium spectrum} & \multicolumn{2}{|c|}{Background spectrum} 
& Difference \\
Energy & Decay & Chain & Intensity & $\varepsilon$ & Position & Rate & Position & Rate & \\
(keV) &  &  & ($\%$) & ($\%$) & (keV) &  c/day & (keV) & c/day & c/day \\
\hline
}
\tabletail{%
\hline
\multicolumn{10}{|c|}{\small\sl\ (continued)}\\  
\hline}
\tablelasttail{\hline}
\topcaption{$\gamma$ lines observed in the indium and background spectra. 
Intensities are expressed as number of $\gamma$'s per nucleus decay. 
Efficiencies refer to the four detectors (in OR) and are evaluated by cubic spline interpolation.
} \label{gammatable}
\begin{supertabular}{|cccc|c|c c|c c|c|}
\hline
238.63 & $^{212}$Pb & $^{232}$Th & 43.3 & 2.55 &  238.34 & 6.03$\pm$1.00 & 238.64 & 6.00$\pm$0.87 & 0.03$\pm$1.33 \\
\hline
242.00 & $^{214}$Pb & $^{238}$U & 7.43 & 2.59 & 241.84 & 2.99$\pm$0.70 & 241.70 & 3.30$\pm$0.63 & -0.30$\pm$0.95 \\
\hline
295.22 & $^{214}$Pb & $^{238}$U &  19.3 & 3.08 & 295.16 & 5.37$\pm$0.82 & 295.23 & 4.47$\pm$0.93 & 0.90$\pm$1.24\\
\hline
300.01  & $^{227}$Th & $^{235}$U &  2.32 & 3.11 & 299.46 & 0.82$\pm$0.63 & 300.73 & 0.69$\pm$0.51 & 0.13$\pm$0.81\\ 
300.07  & $^{212}$Pb & $^{232}$Th & 3.28 & 3.11 & & & & & \\
\hline
338.32 & $^{228}$Ac & $^{232}$Th &  11.27 & 3.27 & 338.33 & 1.20$\pm$0.43 & 337.95 & 1.41$\pm$0.48 & -0.21$\pm$0.65\\
\hline
351.06 & $^{211}$Bi & $^{235}$U  & 12.91 & 3.30 & 351.88 & 8.79$\pm$0.47 & 351.88 & 10.31$\pm$0.96 & -1.51$\pm$1.07 \\
351.93 & $^{214}$Pb & $^{238}$U & 37.6 & 3.30 & &            & &  & \\
\hline
  &  &  &  & 3.32 & 497.48$\pm$0.21 & 0.78$\pm$0.19 & - & $<$0.34 (90$\%$ CL) & - \\
\hline
510.77 & $^{208}$Tl & $^{232}$Th & 22.6 & 3.31 & 510.83 & 4.50$\pm$0.34 & 510.65 & 4.20$\pm$0.51 & 0.30$\pm$0.67\\
510.99 & e$^{+}$ ann & *  & & 3.31     &          &           & &    & \\
\hline
569.70 & $^{211}$Po & $^{235}$U  & 0.5 & 3.25 & 570.11 & 0.71$\pm$0.24 & 570.36 & 0.57$\pm$0.18 & 0.14$\pm$0.30\\
569.70 & $^{207}$Bi & * & 97.7   & 3.25         & & & & &\\
\hline
583.19 & $^{208}$Tl & $^{232}$Th & 84.5 & 3.23 & 583.18 & 1.62$\pm$0.19 & 583.53 & 1.80$\pm$0.27 & -0.18$\pm$0.33\\
\hline
609.31 & $^{214}$Bi & $^{238}$U  &  46.1 & 3.19 & 609.29 & 7.07$\pm$0.28 & 609.27 & 6.96$\pm$0.36 & 0.12$\pm$0.45\\
\hline
661.66 & $^{137}$Cs & *           & 85.1 & 3.13 & 661.76 & 4.34$\pm$0.26 & 661.27 & 4.05$\pm$0.30 & 0.29$\pm$0.40\\
\hline
727.33 & $^{212}$Bi & $^{232}$Th &  6.58 & 3.06 & 727.20 & 0.91$\pm$0.28 & 727.15 & 0.47$\pm$0.33 & 0.44$\pm$0.43 \\
\hline
768.36 & $^{214}$Bi & $^{238}$U  & 4.96 & 3.02 & 767.69 & 0.73$\pm$0.36 & 768.04 & 0.45$\pm$0.18 & 0.28$\pm$0.41\\
\hline
794.95 & $^{228}$Ac & $^{232}$Th & 4.25 & 2.99 & 794.93 & 0.56$\pm$0.21 & - & $<$0.67 (90$\%$ CL)  &  - \\
\hline
803.10 & $^{206}$Pb$^{*}$ & *   &  & 2.98 & 803.11 & 0.71$\pm$0.23 & 803.17 & 1.17$\pm$0.33 & -0.46$\pm$0.40\\
\hline
834.84 & $^{54}$Mn &      *    & 100.0 & 2.95 & 834.82 & 0.73$\pm$0.19 & 835.44 & 0.75$\pm$0.18 & -0.02$\pm$0.26 \\
835.71 & $^{228}$Ac & $^{232}$Th & 1.61 & 2.95 & & & & &\\
\hline
839.04 & $^{214}$Pb & $^{238}$U & 0.59 & 2.95 & 840.89 & 0.50$\pm$0.15 & 839.72 & 0.36$\pm$0.18 & 0.14$\pm$0.23 \\ 
840.37 & $^{228}$Ac & $^{232}$Th & 0.91 & 2.95 &                &               & & &\\
\hline
860.56 & $^{208}$Tl & $^{232}$Th & 12.42 & 2.93 & 859.29 & 0.64$\pm$0.43 & 861.34 & 0.51$\pm$0.39  & 0.13$\pm$0.58\\ 
\hline
911.20 & $^{228}$Ac & $^{232}$Th &  25.8 &  2.87 & 911.30 & 1.79$\pm$0.16 & 911.16 & 1.70$\pm$0.35 & 0.09$\pm$0.38\\ 
\hline
934.10 & $^{214}$Bi & $^{238}$U  & 3.03 & 2.85 & 933.68 & 0.40$\pm$0.14 & 934.53 & 0.21$\pm$0.21 & 0.19$\pm$0.25\\
\hline
964.76 & $^{228}$Ac & $^{232}$Th &  4.99 & 2.82 & 963.60 & 0.41$\pm$0.19 & 964.44 & 0.27$\pm$0.12 & 0.15$\pm$0.23\\ 
\hline
968.97 & $^{228}$Ac & $^{232}$Th &  15.8 & 2.81 & 969.17 & 1.16$\pm$0.26 & 968.95 & 0.84$\pm$0.18 & 0.32$\pm$0.32 \\
\hline
1001.03 & $^{234m}$Pa & $^{238}$U &  0.84 & 2.78 & 1001.00 & 0.35$\pm$0.14 & 1001.43 & 0.30$\pm$0.15 & 0.05$\pm$0.20\\
\hline
1063.66 & $^{207}$Bi & * & 74.5 & 2.73 & 1063.35 & 0.78$\pm$0.14 & 1063.69 & 0.90$\pm$0.15 & -0.12$\pm$0.20\\
\hline
1120.29 & $^{214}$Bi & $^{238}$U & 15.1 & 2.68 & 1120.27 & 1.46$\pm$0.12 & 1119.91 & 1.56$\pm$0.18 & -0.10$\pm$0.22\\
\hline
1124.00 & $^{65}$Zn + Cu K$_{\alpha}$ & * & 50.6 & 2.68 & 1124.09 & 0.89$\pm$0.10 & 1124.02 & 0.96$\pm$0.12 & -0.07$\pm$0.16\\
\hline
1173.24 & $^{60}$Co & *           & 99.97 & 2.65 & 1173.20 & 4.92$\pm$0.32 & 1173.22 & 4.56$\pm$0.48 & 0.36$\pm$0.58 \\
\hline
1238.11 & $^{214}$Bi & $^{238}$U &  5.79 & 2.61 & 1238.12 & 0.72$\pm$0.21 & 1237.81 & 0.84$\pm$0.21 & -0.12$\pm$0.30\\
\hline
1332.50 & $^{60}$Co &      *  & 99.99 & 2.53 & 1332.62 & 4.88$\pm$0.32 & 1332.16 & 4.52$\pm$0.45 & 0.36$\pm$0.55\\
\hline
1377.67 & $^{214}$Bi & $^{238}$U &  4.00 & 2.49 & 1377.53 & 0.44$\pm$0.14 & 1377.59 & 0.39$\pm$0.06 & 0.05$\pm$0.15\\
\hline
1407.98 & $^{214}$Bi & $^{238}$U &  2.15 & 2.46 & 1408.22 & 0.45$\pm$0.24 & 1408.25 & 0.48$\pm$0.30 & -0.03$\pm$0.39\\
\hline
1460.83 & $^{40}$K &       *   &  11.0 & 2.42 & 1460.92 & 18.14$\pm$0.60 & 1460.94 & 18.55$\pm$0.84 & -0.4$\pm$1.0\\
\hline
1509.23 & $^{214}$Bi & $^{238}$U &  2.11 & 2.38 & 1509.02 & 0.24$\pm$0.14 & 1508.86 & 0.24$\pm$0.06  & 0.00$\pm$0.15\\
\hline
1588.19 & $^{228}$Ac & $^{232}$Th &  3.22 & 2.31 & 1588.02 & 0.24$\pm$0.11 & - & $<$ 0.36 (90$\%$ CL) & - \\
\hline
1729.60 & $^{214}$Bi & $^{238}$U & 2.92 & 2.19 & 1729.79 & 0.33$\pm$0.10 & 1729.84 & 0.48$\pm$0.12 & -0.15$\pm$0.16\\
\hline
1764.49 & $^{214}$Bi & $^{238}$U &  15.4 & 2.17 & 1764.49 & 1.58$\pm$0.20 & 1764.00 & 1.78$\pm$0.24 & -0.21$\pm$0.31\\
\hline
1770.24 & $^{207}$Bi & *       &  6.87 & 2.17 & 1769.81 & 0.17$\pm$0.04 & 1770.00 & 0.22$\pm$0.07 & -0.05$\pm$0.08 \\
\hline
1808.65 & $^{26}$Al  &     *   &  99.8 & 2.14 & 1808.14 & 0.38$\pm$0.09 & 1808.44 & 0.30$\pm$0.09 & 0.08$\pm$0.13\\
\hline
1847.42 & $^{214}$Bi & $^{238}$U &  2.11 & 2.11 & 1847.53 & 0.14$\pm$0.07 & 1847.10 & 0.21$\pm$0.09 & -0.07$\pm$0.11\\
\hline
2204.21 & $^{214}$Bi & $^{238}$U &  5.08 & 1.89 & 2203.29 & 0.30$\pm$0.05 & 2203.37 & 0.33$\pm$0.09 & -0.03$\pm$0.10\\
\hline
2447.86 & $^{214}$Bi & $^{238}$U &  1.57 & 1.75 & 2447.86 & 0.19$\pm$0.09 & - & $<$ 0.27 (90$\%$ CL) & - \\
\hline
2614.54 & $^{208}$Tl & $^{232}$Th &  99.0 & 1.66 & 2614.16 & 0.90$\pm$0.14 & 2615.08 & 0.64$\pm$0.17 & 0.26$\pm$0.22\\
\hline
\end{supertabular}
\end{center}
\end{landscape}

\end{document}